\documentclass{article}
\usepackage{amsmath}
\usepackage{graphicx}
\usepackage{subfigure}
\usepackage{authblk}
\usepackage{hyperref}
\usepackage{etoolbox}
\usepackage{lmodern}
\usepackage{abstract}
\usepackage[T1]{fontenc}
\usepackage[utf8]{inputenc}
\usepackage{amsmath}
\usepackage{mathtools}
\usepackage{graphicx}
\usepackage{subfigure}
\usepackage{authblk}
\usepackage{hyperref}
\usepackage{etoolbox}
\usepackage{lmodern}
\usepackage[T1]{fontenc}
\usepackage{braket}
\usepackage[utf8]{inputenc}
\usepackage[font=small,labelfont=bf]{caption} 
\usepackage{geometry}
\usepackage[font=small,labelfont=bf]{caption} 
\usepackage{geometry}
\title{Electronic Structure Calculatins and the Ising Machine}
\author[1]{Rongxin Xia}
\author[1]{Teng Bian}
\affil[1]{Department of Physics, Purdue University, West Lafayette, IN, 47907 USA}
\author[1,2,3,4]{Sabre Kais \thanks{kais@purdue.edu}}
\affil[2]{Department of Chemistry and Birck Nanotechnology Center, Purdue University,
West Lafayette, IN 47907 USA}
\affil[3]{Qatar Environment and Energy Research Institute, HBKU, Doha, Qatar}
\affil[4]{Santa Fe Institute,  1399 Hyde Park Rd, Santa Fe, NM 87501}
\date{}
\begin{document}
\maketitle
\vspace{-8ex}
\begin{abstract}
{\bf The exact solution of Schr\"odinger equation for atoms, molecules and extended systems continues to be a "Holy Grail" problem that the entire field has been striving to solve since its inception. Recently, breakthroughs have been made in the development of quantum annealing and coherent Ising machines capable of simulating hundreds of connected spins interacting with an Ising type Hamiltonian. One of the most vital questions pertaining to these new devices is: 'can these machine be used to perform electronic structure calculations?' Here we discuss the general procedure used by these devices and show that there is an exact mapping between the electronic structure Hamiltonian of the Hydrogen molecule and the Ising Hamiltonian.}
\end{abstract}

Determining solutions to the the Schr\"odinger equation is fundamentally difficult because the dimensionality of the corresponding Hilbert space increases exponentially with the number of particles in the system, which requires a commensurate increase in computational resources. Modern quantum chemistry --- faced with difficulties associated with solving Schr\"odinger equations to an accuracy of $\sim$1 kcal/mole (known as chemical accuracy) --- has largely become an endeavor to find approximate methods. A few products of this effort from the past few decades include: \emph{Ab initio}, Density Functional, Density Matrix, Algebraic, Quantum Monte Carlo and Dimensional Scaling methods\cite{herschbach2012dimensional,iachello1995algebraic,kais_book,szabo1989modern}. However, all methods which have been devised to date face the unsurmountable challenge of computational resource requirements as the calculation is extended either to higher accuracy or to larger systems. Computational complexity in electronic structure\cite{PCCP,Frank} suggests that these restrictions are a result of an inherent difficulty associated with simulating quantum systems.

Electronic structure algorithms developed for quantum computers provide a new promising route to advance the field of electronic structure calculations for large systems\cite{o2015scalable,Seth}. Recently, there has been an attempt at using an adiabatic quantum computing model --- as is implemented on the D-Wave machine --- to perform electronic structure calculations\cite{Alan-SR}. The fundamental concept behind the adiabatic quantum computing (AQC) method is to define a problem Hamiltonian, $H_P$, engineered to have its ground state encode the solution of a corresponding computational problem. The system is then initialized in the ground state of a beginning Hamiltonian, $H_B$, which is easily solved classically; then the system is allowed to evolve adiabatically as: $H(s)=(1-s) H_B+s H_P$ (where $s$ is a time parameter, $s\in[0,1]$).  The adiabatic evolution is governed by the Schr\"odinger equation for the time-dependent Hamiltonian $H(s(t))$.

The largest scale implementation of AQC to hitherto is by D-Wave Systems\cite{Boixo,Rose}. In the case of the D-Wave device, the physical process acting as adiabatic evolution is more broadly called \emph{quantum annealing} (QA). The quantum processors manufactured by D-Wave are essentially a transverse Ising model with tunable local fields and coupling coefficients: $ H = \sum_i\Delta_i \sigma_x^i + \sum_{i}h_i \sigma_z^i + \sum_{i,j}J_{ij}\sigma_z^i \sigma_z^j$, where the parameters $\Delta_i$, $h_i$ and $J_{ij}$ are physically tunable. The qubits (quantum bits) are connected in a specified graph geometry, allowing for the embedding of arbitrary graphs. Zoller and coworker presented a scalable architecture with full connectivity, which can only be implemented with local interactions\cite{Zoller}. The adiabatic evolution is initialized at $ H_B = -h\sum_i \sigma_x^i$ and evolves into the problem Hamiltonian: $\ H_P=\sum_{i}h_i \sigma_z^i + \sum_{i,j}J_{ij} \sigma_z^i \sigma_z^j$.  This equation describes a classical Ising model whose ground state is --- in the worst case --- \textsc{NP}-complete.  Therefore any combinatorial optimization \textsc{NP}-hard problem may be encoded into the parameter assignments, $\{h_i,J_{ij}\}$, of $H_P$ and may exploit the adiabatic evolution under $ H(s)=(1-s) H_B+s H_P$ as a method for reaching the ground state of $H_P$. More recently, an optically-based coherent Ising machine was developed; this machine is capable of finding the ground state of an Ising Hamiltonian that describes a set of hundreds coupled spin-1/2 particles\cite{Ising-1,Ising-2,Ising-3}. These challenging NP-hard problems are characterized by an unlikelihood of a solution defined by a polynomial-time algorithm, therefore solutions cannot be easily found using classical numerical algorithms in a reasonable time for large system sizes ($N$)\cite{Ising-1,Ising-2,Ising-3}. These special purpose machines may help in finding the solutions to some of the hardest problems in computing.

The technical scheme for performing electronic structure calculations on such an Ising-type machine can be summarized in the following four steps: First, write down the electronic structure Hamiltonian via the second quantization method in terms of creation and annihilation fermionic operators; Second, use the Jordan-Wigner or the Bravyi-Kitaev transformation to move from fermionic operators to spin operators\cite{bravyi2002fermionic}; Third, Reduce the Spin Hamiltonian --- which is a k-local in general --- to a 2-local Hamiltonian. Finally, map the 2-local Hamiltonian to an Ising-type Hamiltonian with control errors consistent with chemical accuracy ($\sim 10^{-3}$ Hartree). 

Explicitly, this general procedure begins with a second quantization description of a fermionic system in which $N$ single-particle states can be either empty or occupied by a spineless fermionic particle\cite{mcweeny1969methods,szabo1989modern}. One may then use the tensor product of individual spin orbitals written as $|f_{n}...f_0 \rangle$ to represent states in fermionic systems, where $f_j \in \left\{ 0,1 \right\}$ is the occupation number of orbital $j$. Any interaction within the fermionic system can be expressed in terms of products of the creation and annihilation operators $a_j^{\dagger}$ and $a_j$, for $j \in \left\{0, ..., N\right\}$. Thus, the molecular electronic Hamiltonian can be written as:

\begin{equation}
   \hat H= \sum_{i,j}h_{ij}a_i^\dagger a_j+\frac{1}{2}\sum_{i,j,k,l} h_{ijkl}a_i^\dagger a_j^\dagger a_ka_l.
\end{equation}

The above coefficients $h_{ij}$ and $h_{ijkl}$ are one- and two-electron integrals --- which can be precomputed in a classical fashion --- are an input to the quantum simulation. The next step is to employ a Pauli matrices representation of the creation and annihilation operators. We can then use the Bravyi-Kitaev transformation\cite{bravyi2002fermionic,seeley2012bravyi} as mapping between the operator representation and Pauli matrices, $\left\{\sigma_x,\sigma_y,\sigma_z\right\}$, which obey fermionic commutation relations. The molecular Hamiltonian takes the general form:

\begin{equation}
H=\sum\limits_{i\alpha}h_\alpha^i\sigma_\alpha^i+\sum\limits_{ij\alpha\beta}h_{\alpha\beta}^{ij}\sigma_\alpha^i\sigma_\beta^j+\sum\limits_{ijk\alpha\beta\gamma}h_{\alpha\beta\gamma}^{ijk}\sigma_\alpha^i\sigma_\beta^j\sigma_\gamma^k+...
\end{equation}.

Now, after having developed a k-local spin Hamiltonian (many-body interactions), one should use Hamiltonian gadget theory\cite{Ref-7,Ref-14} to reduce it to 2-local (two-body interactions) spin Hamiltonian; this is a requirement since the proposed experimental systems are typically limited to restricted forms of two-body interactions. Therefore, universal adiabatic quantum computation requires a method for approximating quantum many-body Hamiltonians up to an arbitrary spectral error using at most two-body interactions. Hamiltonian gadgets offer a systematic procedure through which to address this requirement.  Recently, Cao et al\cite{Kais-2015} employed analytical techniques resulting in a reduction of the resource scaling as a function of spectral error for the most commonly used subdivisions, three- to two-body and k-body gadgets.

As an example, herein we present calculations for the hydrogen molecule, $H_{2}$. Using the Bravyi-Kitaev transformation, the spin Hamiltonian for molecular hydrogen in the minimal (STO-6G) basis is given by: \cite{seeley2012bravyi}

\begin{equation}
\label{eqn3_2}
\begin{aligned}
H_{H_2}&=f_0{\bf 1}+f_1\sigma_z^0+f_2\sigma_z^1+f_3\sigma_z^2+f_1\sigma_z^0\sigma_z^1 \\
& +f_4\sigma_z^0\sigma_z^2+f_5\sigma_z^1\sigma_z^3+f_6\sigma_x^0\sigma_z^1\sigma_x^2+f_6\sigma_y^0\sigma_z^1\sigma_y^2 \\
& +f_7\sigma_z^0\sigma_z^1\sigma_z^2+f_4\sigma_z^0\sigma_z^2\sigma_z^3+f_3\sigma_z^1\sigma_z^2\sigma_z^3 \\
& +f_6\sigma_x^0\sigma_z^1\sigma_x^2\sigma_z^3+f_6\sigma_y^0\sigma_z^1\sigma_y^2\sigma_z^3+f_7\sigma_z^0\sigma_z^1\sigma_z^2\sigma_z^3 .
\end{aligned}
\end{equation}

Where --- within the above --- the parameters $\left\{f_i\right\}$ are related to the one- and two-electron integrals for a fixed bond length of the molecule. As showed in \cite{o2015scalable}, we notice that this Hamiltonian acts on off-diagonal elements for only two qubits 0, 2 and qubits 1 and 3 never flip. We can use this symmetry to reduce the Hamiltonian to the following effective Hamiltonian, acting only on two qubits:
\begin{equation}
\begin{aligned}
H_{H_2}&=g_0{\bf 1}+g_1\sigma_z^0+g_2\sigma_z^1+g_3\sigma_z^0\sigma_z^1+g_4\sigma_x^0\sigma_x^1+g_4\sigma_y^0\sigma_y^1\\
&=g_0{\bf 1}+H_0
\end{aligned}
\end{equation}
\begin{equation}
H_0=g_1\sigma_z^0+g_2\sigma_z^1+g_3\sigma_z^0\sigma_z^1+g_4\sigma_x^0\sigma_x^1+g_4\sigma_y^0\sigma_y^1
\end{equation}

By squaring the Hamiltonian $H_0$ and modifying it, one can get a new Ising Hamiltonian:

\begin{equation}
H_1=H_{0}^2+2g_3H_{0}=a_1+a_2(\sigma_z^0+\sigma_z^1)+a_3\sigma_z^0\sigma_z^1
\end{equation}

With:
\begin{equation}
a_1=g_1^2+g_2^2+g_3^2+2g_4^2;\ 
a_2=2(g_1+g_2)g_3;\ 
a_3=2(g_1g_2-g_4^2+g_3^2)\ 
\end{equation}

We have succeeded to develop an exact mapping between the ground state energy of the hydrogen molecule and the Ising-type Hamiltonian. The ground state can be easily obtained and compared with the exact calculations as shown in Figure (1). We include detailed procedure and proof in the Supplementary Material. 
\begin{center}
\includegraphics[height=3in,]{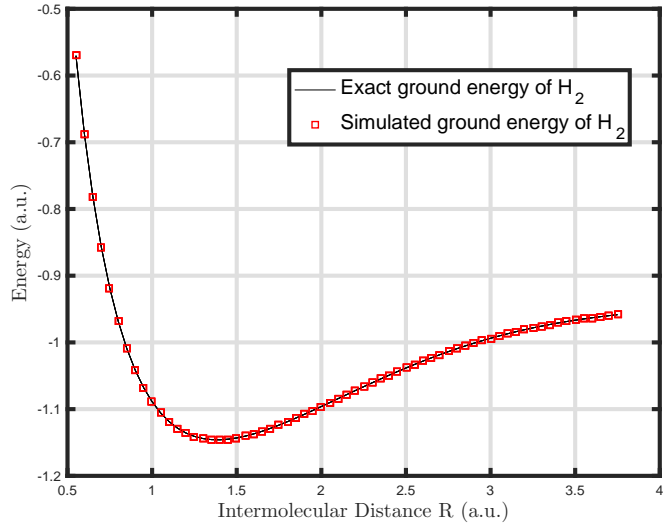}\captionof{figure}{Comparing the numerical results of ground state of the Ising Hamiltonian, Eq.(6) with the exact calculations of the ground state of H$_2$ molecule}
\end{center}

This demonstrates that one can map the electronic ground state energy of a molecular Hamiltonian to an Ising-type Hamiltonian which could easily be implemented on presently available Ising machines. The development of these new Ising machines and the possibility of mapping the electronic structure problem into an Ising-type Hamiltonian may grant efficient ways to obtain exact solutions to the Schr\"odinger equation; this being one of the most daunting computational problem present in both chemistry and physics.

\bibliographystyle{unsrt}
\bibliography{cite12.bbl}
\newpage
\section*{Supplementary Material}
\subsection*{1. Detailed Procedure}
Here we present steps to get the ground state of $H_{H_2}$ by using the new Ising Hamiltonian $H_1$ (Eq.6).

1. If $|g_1|+|g_2|+|g_4|< |g_3|$ start computing by $H_1$ and get the result $Y$. Otherwise increase $|g_3|$ by $|g_1|+|g_2|+|g_4|$ and start computing.

2. Solve equation $x^2+2g_3x=Y$ and get $\sigma_x^1$ and $\sigma_x^2$ ($\sigma_x^1<=\sigma_x^2$). Add $|g_1|+|g_2|+|g_4|$ to $\sigma_x^1$ if added to $g_3$ before (we just assume $g_3>0$.) Compare $\sigma_x^1$ with $g_3-g_1-g_2$ (or $g_3+g_1+g_2$) to get the ground state of $H_0$. Add $g_0$ to get the ground state of $H_{H_2}$.  

\subsection*{2. Theoretical Proof}
We have four eigenvectors of $H_0$:  $|1,1\rangle$ and $|0,0\rangle$ are the two eigenvectors of $H_0$ with eigenvalues $-g_1-g_2+g_3$ and $g_1+g_2+g_3$ respectively (easy to verify).  If $|\psi_3\rangle=a|0,1\rangle+b|1,0\rangle$ is a eigenvector of $H_0$ with eigenvalue $C-g_3$ then $|\psi_4\rangle=-b|0,1\rangle+a|1,0\rangle$ is also eigenvector of $H_0$ with eigenvalue $-C-g_3$. To show this results: Since $\sigma_z^0\sigma_z^1|\psi_1\rangle=-|\psi_1\rangle$ for any $a$ and $b$, we can just consider $H_0=g_1\sigma_z^0+g_2\sigma_z^1+g_4\sigma_x^0\sigma_x^1+g_4\sigma_y^0\sigma_y^1$. For $|\psi_3\rangle$ we have:

\begin{equation}
	g_1a-g_2a+2g_4b=Ca;   \;   -g_1b+g_2b+2g_4a=Cb
\end{equation}

By replacing $a$ with $-b$ and $b$ with $a$ we have:

\begin{equation}
	-g_1b+g_2b+2g_4a=(-C)(-b);  \; 
	g_1a+g2a+-2g_4a=-Ca
\end{equation}

Thus $|\psi_4\rangle=-b|0,1\rangle+a|1,0\rangle$ is also eigenvector of $H_0$ with eigenvalue $-C-g_3$. Now, by changing $g_3$ we can always make sure we can get the ground eigenvalue of $H_0$ by using the eigenvalue of $H_1$. This can be  proved as before, the four eigenvalue of $H$ is $g_1+g_2+g_3, -g_1-g_2+g_3, -g_3-C$ and $-g_3+C$.  Here we set $x_0$ to be the eigenvalue of $H_0$ which corresponds to the ground eigenvalue of $H_1$. Thus $x_0$ must make the function $f(x)=x^2+2g_3x$ to be the smallest, which means $|x_0+g_3|$ should be the smallest. Thus when $|-C-g_3+g_3|<min(|g_1+g_2+g_3+g_3|,|-g_1-g_2+g_3+g_3|$ or $|g_1|+|g_2|+|g_4|< g_3$ (here we just assume $g_3 > 0$), $x_0=-g_3-C$ and we can get $x_0$ by the ground eigenvalue of $H_1$. when $|g_1|+|g_2|+|g_4| > |g_3|$ we can increase $g_3$ with $\Delta$ to make sure $x_0=-g_3-\Delta-C$ and after we get $x_0$ we can add $\Delta$ to get eigenvalue $-g_3-C$. Finally, by comparing $-g_3-C$ and $g_3-g_1-g_2$ (or $g_3+g_1+g_2$) one get the ground eigenvalue of $H_0$.

\subsection*{3. Electronic structure problem }

We treat Hydrogen molecule in a minimal basis STO-6G. Considering spin functions, the four molecular spin orbitals in $H_2$ is:
\begin{align}
	\ket{\chi_1} = \ket{\phi_g}\ket{\alpha} = \frac{ \ket{\phi_{1s}}_1+\ket{\phi_{1s}}_2}{\sqrt{2(1+S)}}\ket{\alpha}
\end{align}

\begin{align}
	\ket{\chi_2} = \ket{\phi_g}\ket{\beta} = \frac{ \ket{\phi_{1s}}_1+\ket{\phi_{1s}}_2}{\sqrt{2(1+S)}}\ket{\beta}
\end{align}

\begin{align}
	\ket{\chi_3} = \ket{\phi_u}\ket{\alpha} = \frac{ \ket{\phi_{1s}}_1-\ket{\phi_{1s}}_2}{\sqrt{2(1-S)}}\ket{\alpha}
\end{align}

\begin{align}
	\ket{\chi_4} = \ket{\phi_u}\ket{\beta} = \frac{ \ket{\phi_{1s}}_1-\ket{\phi_{1s}}_2}{\sqrt{2(1-S)}}\ket{\beta}
\end{align}

where $\ket{\phi_{1s}}_1$ and $\ket{\phi_{1s}}_2$ are spatial-function for two atoms respectively, and $S ={}_{\raisebox{1.535pt}{\scalebox{0.74}{1}}}{\braket{\phi_{1s}|\phi_{1s}}}_2$ where $S$ is the overlap integral.

\begin{align}
	h_{ij} = \int{d\vec{r}\chi_i^*(\vec{r})(-\frac{1}{2}\nabla-\frac{Z}{r})\chi_j(\vec{r})}
\end{align}

\begin{align}
	h_{ijkl} = \int{d\vec{r_1}d\vec{r_2}\chi_{i}^*(\vec{r_1})\chi_{j}^*(\vec{r_2})\frac{1}{r_{12}}\chi_{k}(\vec{r_2})\chi_{l}(\vec{r_1})}
\end{align}

Thus we can write second quantization Hamiltonian of $H_2$:
\begin{equation}
\begin{aligned}
	H_{H_2} &= h_{00}a_0^\dagger a_0 + h_{11}a_1^\dagger a_1+ h_{22}a_2^\dagger a_2 + h_{33}a_3^\dagger a_3+
	h_{0110}a_0^\dagger a_1^\dagger a_1 a_0 +h_{2332}a_2^\dagger a_3^\dagger a_3 a_2  + h_{0330}a_0^\dagger a_3^\dagger a_3 a_0\\
	&+ h_{1221} a_1^\dagger a_2^\dagger a_2 a_1  + (h_{0220} - h_{0202})a_0^\dagger a_2^\dagger a_2 a_0
	+ (h_{1331}-h_{1313})a_1^\dagger a_3^\dagger a_3 a_1 \\
	&+h_{0132}(a_0^\dagger a_1^\dagger a_3 a_2 + a_2^\dagger a_3^\dagger a_1 a_0)+
	h_{0312}(a_0^\dagger a_3^\dagger a_1 a_2 + a_2^\dagger a_1^\dagger a_3 a_0)
\end{aligned}
\end{equation}

By using Bravyi-Kitaev transformation, we have:
\begin{equation}
\label{eqn3_2}
\begin{aligned}
&a_0^{\dagger}=\frac{1}{2}\sigma_x^3\sigma_x^1(\sigma_x^0- i\sigma_y^0)\quad
a_0=\frac{1}{2}\sigma_x^3\sigma_x^1(\sigma_x^0+i\sigma_y^0)\quad
a_1^{\dagger}=\frac{1}{2}(\sigma_x^3\sigma_x^1\sigma_z^0- i\sigma_x^3\sigma_y^1)\quad
a_1=\frac{1}{2}(\sigma_x^3\sigma_x^1\sigma_z^0+i\sigma_x^3\sigma_y^1)\\
&a_2^{\dagger}=\frac{1}{2}\sigma_x^3(\sigma_x^2- i\sigma_y^2)\sigma_z^1\quad
a_2=\frac{1}{2}\sigma_x^3(\sigma_x^2+ i\sigma_y^2)\sigma_z^1\quad
a_3^{\dagger}=\frac{1}{2}(\sigma_x^3\sigma_z^2\sigma_z^1- i\sigma_y^3)\quad
a_3=\frac{1}{2}(\sigma_x^3\sigma_z^2\sigma_z^1+ i\sigma_y^3)
\end{aligned}
\end{equation}
Thus, the Hamiltonian of $H_2$ takes the form:
\begin{equation}
\label{eqn3_2}
\begin{aligned}
H_{H_2}&=f_0{\bf 1}+f_1\sigma_z^0+f_2\sigma_z^1+f_3\sigma_z^2+f_1\sigma_z^0\sigma_z^1 \\
   & +f_4\sigma_z^0\sigma_z^2+f_5\sigma_z^1\sigma_z^3+f_6\sigma_x^0\sigma_z^1\sigma_x^2+f_6\sigma_y^0\sigma_z^1\sigma_y^2 \\
   & +f_7\sigma_z^0\sigma_z^1\sigma_z^2+f_4\sigma_z^0\sigma_z^2\sigma_z^3+f_3\sigma_z^1\sigma_z^2\sigma_z^3 \\
   & +f_6\sigma_x^0\sigma_z^1\sigma_x^2\sigma_z^3+f_6\sigma_y^0\sigma_z^1\sigma_y^2\sigma_z^3+f_7\sigma_z^0\sigma_z^1\sigma_z^2\sigma_z^3      
\end{aligned}
\end{equation}

We notice that qubits 1,3 never flip. We can use this symmetry to reduce the Hamiltonian to the following form which just acts on only two qubits:

\begin{equation}
H_{H_2}=g_0{\bf 1}+g_1\sigma_z^0+g_2\sigma_z^1+g_3\sigma_z^0\sigma_z^1+g_4\sigma_x^0\sigma_x^1+g_4\sigma_y^0\sigma_y^1
\end{equation}

\begin{equation}
g_0=f_0\ g_1=2f_1\ g_2=2f_3\ g_3=2(f_4+f_7)\ g_4=2f_6
\end{equation}

Where $\left\{g_i\right\}$ depends on the fixed bond length of the molecule. In Table I, we present the numerical values of $\left\{g_i\right\}$ in the minimal basis STO-6G.

\begin {table}
\caption {Comparing the Exact Ground Energy (a.u.) Eq.(4), and the Simulated Ground Energy (a.u.) Eq.(6) as a function of the intermolecule dustance R (a.u.)} \label{tab:title} 
\begin{center}
\begin{tabular}{l*{7}{c}} R & $g_0$ & $g_1$ & $g_2$ & $g_3$ & $g_4$ & Exact & Simulated  \\ \hline 0.6 & 1.5943 & 0.5132 & -1.1008 & 0.6598 & 0.0809 & -0.5703 & -0.5703\\0.65 & 1.4193 & 0.5009 & -1.0366 & 0.6548 & 0.0813 & -0.6877 & -0.6877\\0.7 & 1.2668 & 0.4887 & -0.9767 & 0.6496 & 0.0818 & -0.7817 & -0.7817\\0.75 & 1.1329 & 0.4767 & -0.9208 & 0.6444 & 0.0824 & -0.8575 & -0.8575\\0.8 & 1.0144 & 0.465 & -0.8685 & 0.639 & 0.0829 & -0.9188 & -0.9188\\0.85 & 0.909 & 0.4535 & -0.8197 & 0.6336 & 0.0835 & -0.9685 & -0.9685\\0.9 & 0.8146 & 0.4422 & -0.774 & 0.6282 & 0.084 & -1.0088 & -1.0088\\0.95 & 0.7297 & 0.4313 & -0.7312 & 0.6227 & 0.0846 & -1.0415 & -1.0415\\1.0 & 0.6531 & 0.4207 & -0.691 & 0.6172 & 0.0852 & -1.0678 & -1.0678\\1.05 & 0.5836 & 0.4103 & -0.6533 & 0.6117 & 0.0859 & -1.0889 & -1.0889\\1.1 & 0.5204 & 0.4003 & -0.6178 & 0.6061 & 0.0865 & -1.1056 & -1.1056\\1.15 & 0.4626 & 0.3906 & -0.5843 & 0.6006 & 0.0872 & -1.1186 & -1.1186\\1.2 & 0.4098 & 0.3811 & -0.5528 & 0.5951 & 0.0879 & -1.1285 & -1.1285\\1.25 & 0.3613 & 0.372 & -0.523 & 0.5897 & 0.0886 & -1.1358 & -1.1358\\1.3 & 0.3167 & 0.3631 & -0.4949 & 0.5842 & 0.0893 & -1.1409 & -1.1409\\1.35 & 0.2755 & 0.3546 & -0.4683 & 0.5788 & 0.09 & -1.1441 & -1.1441\\1.4 & 0.2376 & 0.3463 & -0.4431 & 0.5734 & 0.0907 & -1.1457 & -1.1457\\1.45 & 0.2024 & 0.3383 & -0.4192 & 0.5681 & 0.0915 & -1.1459 & -1.1459\\1.5 & 0.1699 & 0.3305 & -0.3966 & 0.5628 & 0.0922 & -1.1450 & -1.1450\\1.55 & 0.1397 & 0.323 & -0.3751 & 0.5575 & 0.093 & -1.1432 & -1.1432\\1.6 & 0.1116 & 0.3157 & -0.3548 & 0.5524 & 0.0938 & -1.1405 & -1.1405\\1.65 & 0.0855 & 0.3087 & -0.3354 & 0.5472 & 0.0946 & -1.1371 & -1.1371\\1.7 & 0.0612 & 0.3018 & -0.317 & 0.5422 & 0.0954 & -1.1332 & -1.1332\\1.75 & 0.0385 & 0.2952 & -0.2995 & 0.5371 & 0.0962 & -1.1287 & -1.1287\\1.8 & 0.0173 & 0.2888 & -0.2829 & 0.5322 & 0.097 & -1.1239 & -1.1239\\1.85 & -0.0023 & 0.2826 & -0.267 & 0.5273 & 0.0978 & -1.1187 & -1.1187\\1.9 & -0.0208 & 0.2766 & -0.252 & 0.5225 & 0.0987 & -1.1133 & -1.1133\\1.95 & -0.0381 & 0.2707 & -0.2376 & 0.5177 & 0.0995 & -1.1077 & -1.1077\\2.0 & -0.0543 & 0.2651 & -0.2238 & 0.513 & 0.1004 & -1.1019 & -1.1019\\2.05 & -0.0694 & 0.2596 & -0.2108 & 0.5084 & 0.1012 & -1.0961 & -1.0961\\2.1 & -0.0837 & 0.2542 & -0.1983 & 0.5039 & 0.1021 & -1.0901 & -1.0901\\2.15 & -0.097 & 0.249 & -0.1863 & 0.4994 & 0.103 & -1.0842 & -1.0842\\2.2 & -0.1095 & 0.244 & -0.1749 & 0.495 & 0.1038 & -1.0782 & -1.0782\\2.25 & -0.1213 & 0.2391 & -0.164 & 0.4906 & 0.1047 & -1.0723 & -1.0723\\2.3 & -0.1323 & 0.2343 & -0.1536 & 0.4864 & 0.1056 & -1.0664 & -1.0664\\2.35 & -0.1427 & 0.2297 & -0.1436 & 0.4822 & 0.1064 & -1.0605 & -1.0605\\2.4 & -0.1524 & 0.2252 & -0.1341 & 0.478 & 0.1073 & -1.0548 & -1.0548\\2.45 & -0.1616 & 0.2208 & -0.125 & 0.474 & 0.1082 & -1.0492 & -1.0492\\2.5 & -0.1703 & 0.2165 & -0.1162 & 0.47 & 0.109 & -1.0437 & -1.0437\\2.55 & -0.1784 & 0.2124 & -0.1079 & 0.466 & 0.1099 & -1.0383 & -1.0383\\2.6 & -0.1861 & 0.2083 & -0.0999 & 0.4622 & 0.1108 & -1.0331 & -1.0331\\2.65 & -0.1933 & 0.2044 & -0.0922 & 0.4584 & 0.1117 & -1.0280 & -1.0280\\2.7 & -0.2001 & 0.2006 & -0.0848 & 0.4547 & 0.1125 & -1.0231 & -1.0231\\2.75 & -0.2064 & 0.1968 & -0.0778 & 0.451 & 0.1134 & -1.0184 & -1.0184\\2.8 & -0.2125 & 0.1932 & -0.071 & 0.4475 & 0.1142 & -1.0139 & -1.0139\\2.85 & -0.2182 & 0.1897 & -0.0646 & 0.4439 & 0.1151 & -1.0095 & -1.0095\\2.9 & -0.2235 & 0.1862 & -0.0584 & 0.4405 & 0.1159 & -1.0053 & -1.0053\\ 2.95 & -0.2286 & 0.1829 & -0.0524 & 0.4371 & 0.1168 & -1.0013 & -1.0013\\3.0 & -0.2333 & 0.1796 & -0.0467 & 0.4338 & 0.1176 & -0.9974 & -0.9974\\3.05 & -0.2378 & 0.1764 & -0.0413 & 0.4305 & 0.1184 & -0.9938 & -0.9938\\3.1 & -0.2421 & 0.1733 & -0.036 & 0.4273 & 0.1193 & -0.9903 & -0.9903\\\end{tabular}
\end{center}
\end{table}
\end{document}